\documentclass[]{aastex701}

\usepackage{amsmath}

\begin{document}

\title[Article Title]{A unified picture of swirl-driven coronal heating: magnetic energy supply and dissipation}

\author[orcid=0000-0003-1134-2770, gname=Hidetaka, sname=Kuniyoshi]{Hidetaka Kuniyoshi}
\affiliation{Department of Mathematics, Physics and Electrical Engineering, Northumbria University, UK}
\email[show]{hidetaka.kuniyoshi@northumbria.ac.uk}  

\author[orcid=0000-0001-7891-3916,gname=Shinsuke, sname=Imada]{Shinsuke Imada} 
\affiliation{Department of Earth and Planetary Science, The University of Tokyo, Japan}
\email{imada@eps.s.u-tokyo.ac.jp}

\author[orcid=0000-0001-5457-4999, gname=Takaaki,sname=Yokoyama]{Takaaki Yokoyama}
\affiliation{Astronomical Observatory, Kyoto University, Japan}
\email{yokoyama.takaaki.2a@kyoto-u.ac.jp}

\begin{abstract}

The coronal heating problem is one of the most critical challenges in solar physics.
Recent observations have revealed that small-scale swirls are ubiquitous in the photosphere and chromosphere, suggesting that they may play a significant role in transferring magnetic energy into the corona. However, the overall contribution of swirls to the total magnetic energy supply and subsequent coronal heating remains uncertain.
To address this, we perform statistical analyses of simulated swirls using a three-dimensional radiative magnetohydrodynamic simulation extending from the convection zone to the corona in the quiet Sun. Our results reveal that swirls account for approximately half of the total magnetic energy. Furthermore, they strongly suggest that swirls can trigger coronal heating events through magnetic reconnection. The occurrence frequency of these events follows a power-law-like distribution, consistent with observations of coronal heating signatures known as ``nanoflares'', indicating that swirls are promising candidates as their drivers.

\end{abstract}



\section{Introduction}\label{sec:introduction}

The solar coronal heating problem remains a major challenge in astrophysics.
Although it is widely accepted that the corona is magnetically heated \citep[see][]{Reiners_2012_LRSP}, the mechanisms by which magnetic energy is supplied and converted into heat remain unclear \citep{Klimchuk_2006_SoPh}.
Convective motions in the photosphere interact with magnetic fields, generating Poynting flux that propagates upward into the corona.
Many studies suggest that horizontal plasma motions interacting with vertical magnetic fields are the dominant source \citep[e.g.,][]{Klimchuk_2006_SoPh, Shelyag_2012_ApJ}, while the role of vertical motions on inclined fields (i.e., flux emergence) remains under debate \citep{Shoda_2023_ApJ}.
Traditionally, horizontally driven Poynting flux is attributed to granular motions that randomly displace magnetic flux tubes as a whole \citep{Berger_1998_ApJ}, with the resulting energy transported via kink Alfv{\'e}nic waves \citep{Cranmer_2005_ApJS}.
Although observations have confirmed the ubiquity of kink Alfvénic waves throughout the corona \citep{Morton_2025_ApJ}, it remains debated whether they provide sufficient Poynting flux to compensate for coronal energy losses \citep{Thurgood_2014_ApJ}.

An alternative magnetic energy transport mechanism has been proposed more recently, inspired by the discovery of swirling motions in quiet Sun photosphere and chromosphere \citep{Bonet_2008_ApJ, Wedemeyer_2009_AA, Dakanalis_2022_AA}.
These swirls, originating from convective downdrafts in intergranular lanes, are considered to twist magnetic field lines within photospheric flux tubes, exciting torsional Alfvénic waves \citep{Wedemeyer_2012_Natur, Liu_2019_NatCo}.
\added{It is worth noting that torsional Alfvénic waves can also be driven by other mechanisms, such as interchange magnetic reconnection between unipolar and bipolar magnetic fields \citep{Cranmer_2018_ApJ, Sterling_2020_ApJ}.}
However, the contribution of torsional Alfvénic waves to coronal energy input remains unclear, largely due to limited observational evidence—likely a consequence of insufficient spatial resolution.
Numerical simulations, on the other hand, have shown that individual swirls can enhance the coronal Poynting flux by a factor of $3$--$4$ compared to non-swirling regions \citep{Kuniyoshi_2023_ApJ, Silva_2024_ApJ}.
These results suggest that swirl-driven torsional Alfv{\'e}nic waves may transport significantly more magnetic energy into the corona than granulation-driven kink Alfv{\'e}nic waves, potentially leading to more efficient heating.
Nonetheless, comprehensive statistical studies are needed to validate this scenario.

The connection between magnetic energy supply and dissipation is poorly understood. 
\added{Analytical and numerical studies suggest that non-potential magnetic energy is stored through field line braiding driven by photospheric motions, and subsequently released via magnetic reconnection \citep{Parker_1983_ApJ, Galsgaard_1996_JGR}.
Later simulations have identified granulation as the primary driver of this process \citep{Gudiksen_2005_ApJ, Rempel_2017_ApJ}.}
However, recent modeling efforts estimate that such granulation-driven heating can release only $\sim10^{21} \ \rm erg$ \citep{Einaudi_2021_ApJ, Judge_2023_ApJ}, which is several orders of magnitude lower than the most commonly observed energy range of $10^{23}$--$10^{24} \ \rm erg$, known as the nanoflare regime \citep{Chitta_2021_AA}. 
Magnetic flux cancellation may contribute \citep{Panesar_2021_ApJ}, but observations show brightenings often lack magnetic bipoles \citep{Nelson_2024_aa}, arguing against it as a dominant driver. In contrast, coronal brightenings above chromospheric swirls \citep{Wedemeyer_2012_Natur} and supporting simulations \citep{Kuniyoshi_2024_ApJ} suggest a potential causal link. While these findings point to swirls as potential drivers of nanoflare-like coronal heating events, its ability to reproduce observed statistical properties, such as power-law energy distributions \citep[e.g.,][]{Kawai_2021_ApJ, Upendran_2022_ApJ}, remains to be confirmed.

In this study, we examine the role of swirls in coronal heating, focusing on (1) quantifying the magnetic energy transported into the corona by swirl-driven torsional Alfv{\'e}nic waves, and (2) evaluating whether swirl-driven heating can reproduce the statistical properties of coronal heating events. 
We use a three-dimensional radiative magnetohydrodynamic (MHD) simulation spanning from the upper convection zone to the corona. This approach captures the dynamic coupling across layers \citep{Hansteen_2015_ApJ, Rempel_2017_ApJ}, enabling self-consistent modeling of energy generation, transport, and dissipation for the coronal heating.

\section{Methods}
\label{sec:simulation_setup}

We perform a three-dimensional MHD simulation using the RAdiation Magnetohydrodynamics Extensive Numerical Solver (RAMENS) code \citep{Iijima_2016_PhDT} to model a gravitationally stratified solar atmosphere.
The code solves the compressible MHD equations with Spitzer-H{\"a}rm thermal conduction \citep{Spitzer_1953_PhRv} and \added{optically thick and thin radiation}.
\added{For the optically thick part, corresponding to upper convection zone through lower chromosphere, we solve the gray-approximated local thermodynamic equilibrium (LTE) radiative transfer equation.
For the optically thin part, corresponding to upper chromosphere through corona, we calculate the radiative heating rate using the loss function retrieved from the CHIANTI atomic data base \citep{Dere_1997_AAS, Landi_2012_ApJ} and its chromospheric extension \citep{Goodman_2012_ApJ}.}
\added{Following the approach associated with the OPAL oppacities \citep{Rogers_1996_ApJ}, the equation of state includes contributions from the six most abundant elements in the solar atmosphere (H, He, C, N, O, and Ne).}
Explicit viscosity and resistivity are not included to minimize dependence on spatial resolution.
As a result, all heating arises from numerical dissipation, though it is physically meaningful, since the underlying small-scale, i.e., dissipative structures originate from physical processes such as magnetic field line twisting.
We estimate the heating rate via the numerical dissipation rate ($Q_{\rm a}$; see Appendix).
Further details of the simulation setup are provided in \citet{Iijima_2016_PhDT}.

Following the methodology of \citet{Breu_2022_AA}, we self-consistently simulate a full coronal loop extending from the upper convection zone through the photosphere and chromosphere into the corona, neglecting its curvature.
We model the loop as semi-circular, accounting only for the component of gravitational acceleration along the loop axis.
The simulation domain spans $14 \times 14 \ \rm Mm^2$ horizontally ($x$ and $y$) and $28 \ \rm Mm$ vertically ($z$), covering $-2 \ \rm Mm \le z \le 26 \ \rm Mm$.
Grid size is $60 \ \rm km$ in the horizontal and vertical direction.
The top ($z = 26 \ \rm Mm$) and bottom ($z = -2 \ \rm Mm$) boundaries lie within the convection zone and extend $2 \ \rm Mm$ below the optical depth unity surfaces, located at $z = 0 \ \rm Mm$ and $z = 24 \ \rm Mm$, respectively.
Periodic boundary conditions are applied horizontally.
At the top and bottom boundaries, open boundary conditions allow outflows, while the entropy of inflows is fixed to emulate convective energy transport from the deep convection zone.

We assume an initially plane-parallel, gravitationally stratified atmosphere in the vertical direction.
Initial profiles for mass density ($\rho$) and internal energy density ($e_{\rm int}$) in the convection zone are taken from Model S \citep{ChristensenDalsgaard_1996_Sci}.
Above the surface, the atmosphere is extrapolated under hydrostatic equilibrium to establish a coronal temperature of $1 \ \rm MK$.
The initial velocity is set to zero ($v_x = v_y = v_z = 0$), and the magnetic field is uniformly vertical ($B_x = B_y = 0$, $B_z = 10 \ \rm G$), consistent with typical quiet-Sun coronal values \citep{Klimchuk_2006_SoPh}.

The system is evolved for $10800 \ \rm s$ until convection reaches a quasi-steady state, where the enthalpy flux through the top and bottom boundaries approximately balances the radiative flux at the surfaces.
During this phase, an artificial thermal conductive flux is imposed at the mid-plane ($z = 12 \ \rm Mm$) to maintain a coronal temperature of $1 \ \rm MK$.
The conductive flux is then removed, and the simulation continues for another $10800 \ \rm s$, allowing the corona to evolve self-consistently.
We analyze the final $6600 \ \rm s$ ($15000 \ \rm s \le t \le 21600  \ \rm s$), with snapshots taken every $10 \ \rm s$.

\begin{figure*}[t!]
  \centering
  \includegraphics[width=15cm]{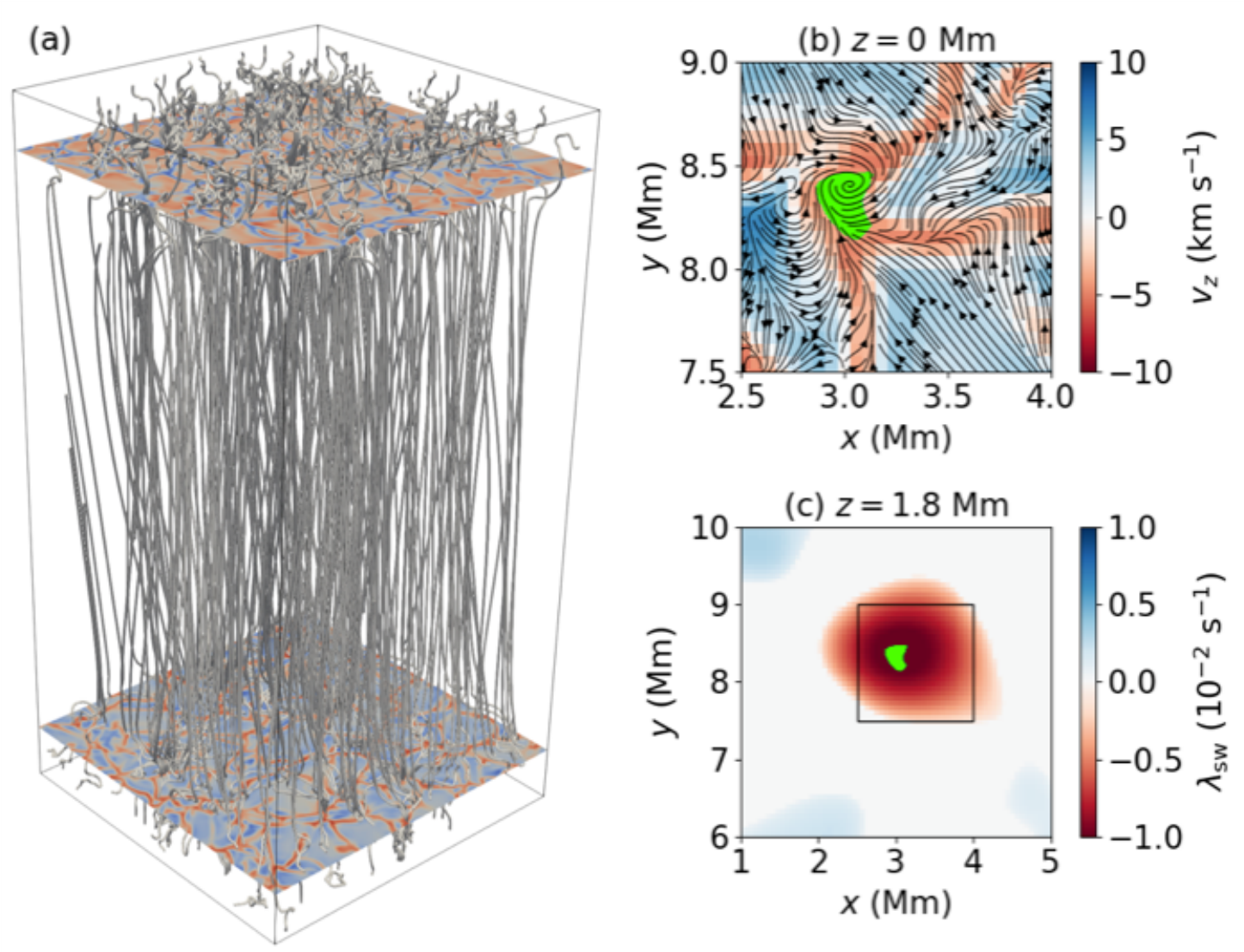} 
  \caption{
  (a) 3D magnetic field lines at $t=17500 \ \rm s$ with vertical velocity ($v_z$) maps at $z=0$ and $z=24 \ \rm Mm$.
(b) Vertical velocity ($v_z$) at $z=0 \ \rm Mm$ in color, with streamlines of horizontal velocity ($v_x, v_y$). Green contour marks a magnetic flux concentration (MC), defined by $|B_z| \ge 500 \ \rm G$.
(c) Swirling strength ($\lambda_{\rm sw}$) at $z=1.8 \ \rm Mm$ with horizontal velocity arrows. Green patch: $B_z \ge 500 \ \rm G$ at $z=0 \ \rm Mm$. Rectangle indicates panel (b) region.
  }
  \label{fig:overview}
\end{figure*}

\section{Results}

\subsection{Swirl properties}
\label{sec:swirl_properties}

Figure~\ref{fig:overview}a shows a three-dimensional visualization of the magnetic field lines from a snapshot of our simulation. 
The coronal magnetic field lines are rooted at both surfaces ($z=0 \ \rm Mm$ and $z=24 \ \rm Mm$), exhibiting a magnetic field geometry that resembles coronal loops rooted in quiet Sun network regions.
A large fraction of the magnetic flux at the surface is concentrated within the downflowing intergranular lanes, forming magnetic flux concetrations (MCs), an example of which is illustrated in Figure~\ref{fig:overview}b.
This panel also displays the horizontal velocity field at the surface, revealing swirling motions within the intergranular lanes. 
Such photospheric swirls are ubiquitously generated within these lanes, creating conditions that allow them to propagate into the upper atmosphere along magnetic field lines as torsional Alfv{\'e}nic wave pulses.
\added{It is important to note that other drivers, such as photospheric or chromospheric reconnection \citep{Sterling_2020_ApJ} and mode coupling \citep{Cally_2008_SoPh}, may also contribute to the generation of torsional Alfvénic waves.
However, directly identifying their origin remains highly challenging due to the complexity of the intervening magnetic field structures.
In contrast, the magnetic topology between the chromosphere and corona is considerably simpler.
Accordingly, the present study focuses on swirls in the chromosphere and transition region, which more directly influence the coronal energy supply.}

We first examine swirls in the chromospheric planes at $z=1.8 \ \rm Mm$ and $z=22.2 \ \rm Mm$, hereafter CH1 and CH2.
Swirls are identified by computing the swirling strength $\lambda_{\rm sw}$ in each plane, a method widely used for swirl detection \citep{Chong_1990_PhFlA, Kato_2017_AA}.
\added{$\lambda_{\rm sw}$ is derived from the gradient tensor of the velocity field projected onto each plane.
In regions of swirling motion, the velocity gradient tensor has complex conjugate eigenvalues.
The magnitude of $\lambda_{\rm sw}$ corresponds to twice the absolute value of the imaginary part of these eigenvalues, making $\lambda_{\rm sw}$ equivalent to the vorticity in the case of pure rotation without shearing motions.
The direction of rotation can be inferred from the associated eigenvectors.
For a more detailed explanation, see \citet{Kato_2017_AA} and \citet{CaniveteCuissa_2020_AA}.}

Due to its dependence on velocity gradients, $\lambda_{\rm sw}$ tends to highlight the smallest-scale swirls.
To mitigate this, we follow \citet{Yadav_2020_ApJ} and apply a Gaussian filter to the velocity field (FWHM = $0.6 \ \rm Mm$), matching the lower size limit of observed chromospheric swirls \citep{Park_2016_AA}.
Figure~\ref{fig:overview}c shows an example of $\lambda_{\rm sw}$ computed from this smoothed field.
Swirl detection criteria are defined as follows:

\begin{enumerate}

    \item At each snapshot, we search the regions of $|\lambda_{\rm sw}| > \mu + n_{\rm min} \sigma$, where $\mu$ and $\sigma$ represent the mean and the standard deviation of $|\lambda_{\rm sw}|$ at each analyzed layer, and $n_{\rm min}$ is an arbitrary constant, respectively. 
    
    \item We track the individual swirls detected by this criterion in both space and time by using the density-based spatial clustering of applications with noise (DBSCAN) method \citep{Ester_1996_kddm}.

    \item We exclude swirls for which the maximum values of $|\lambda_{\rm sw}|$ during their lifetime are below $\mu + n_{\rm max} \sigma$, where $n_{\rm max}$ is another arbitrary constant satisfying $n_{\rm max} \ge n_{\rm min}$.
    The use of this second threshold, $n_{\rm max}$, ensures the detection of spatio-temporally coherent swirls, while their birth and death times are determined by the first threshold, $n_{\rm min}$.

    \item Finally, to ensure the detection of temporally coherent swirls, we exclude the swirls with lifetimes shorter than $1.5 \ \rm min$, which is consistent with the lower limit derived from previous chromospheric observation \citep{Dakanalis_2022_AA}.
    
\end{enumerate}


We perform a parameter survey varying both $n_{\rm min}$ and $n_{\rm max}$ from 1 to 6.
For each case, we compute the occurrence rate, time-averaged surface density, lifetime, and diameter of the swirls.
The occurrence rate is defined as the total number of swirls divided by the area ($14 \times 14 \ \rm Mm^2$) and duration ($6600 \ \rm s$).
Surface density is the temporal mean of the swirl count per unit area.
Assuming circular swirls, the diameter $d_{\rm sw}$ is estimated via $\pi (d_{\rm sw}/2)^2 = a_{\rm sw}$, where $a_{\rm sw}$ is the swirl area.
As CH1 and CH2 yield nearly identical statistics, we report their mean values.
Among all cases, $n_{\rm min} = 2$ shows the least dispersion and best matches previous chromospheric observations \citep{Dakanalis_2022_AA}.
For instance, 358 swirls (189 in CH1, 169 in CH2) are detected when $n_{\rm min} = n_{\rm max} = 2$, closely matching observational results.
The corresponding values (occurrence rate: $0.008 \ \rm Mm^{-2} \ min^{-1}$, surface density: $0.03 \ \rm Mm^{-2}$, lifetime: $3.4 \ \rm min$, and diameter: $1.4 \ \rm Mm$) are consistent with observed values ($0.01 \ \rm Mm^{-2} \ min^{-1}$, $0.08 \ \rm Mm^{-2}$, $8.5 \ \rm Mm$, $2.6 \ \rm min$) within a factor of $\sim 2$ \citep{Dakanalis_2022_AA}. This agreement supports the validity of our model.

\subsection{Energy transfer into the corona by swirls}\label{sec:swirl_energy_transfer}

\begin{figure*}[htbp]
  \centering
  \includegraphics[width=15cm]{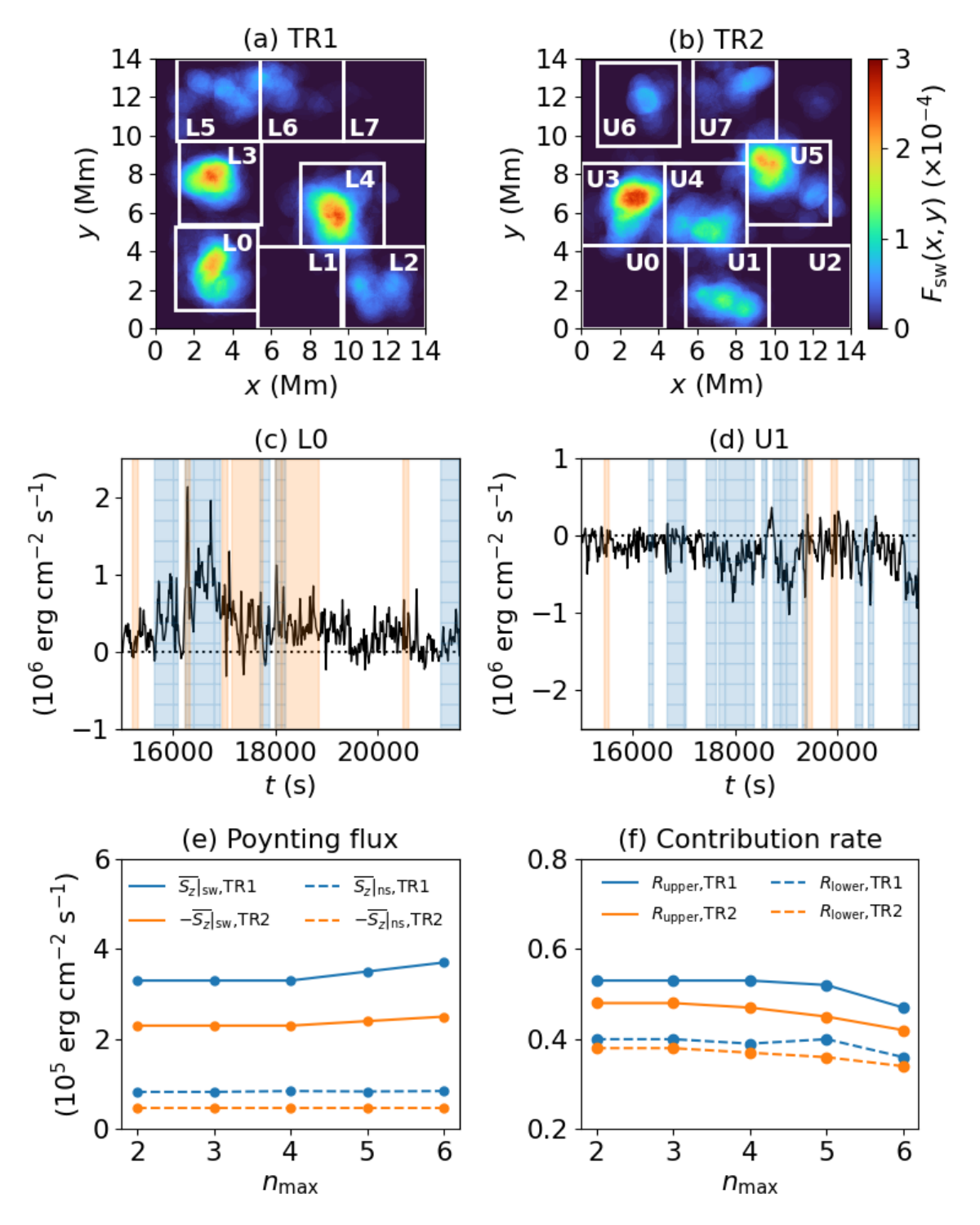} 
  \caption{
  (a, b) Time-integrated swirl distributions $F_{\rm sw}(x,y)$ at TR1 ($z=3.6$ Mm) and TR2 ($z=20.4$ Mm) for $n_{\rm min}=2$, $n_{\rm max}=6$. Color indicates normalized swirl counts.
(c, d) Temporal evolution of $xy$-integrated Poynting flux in mini-squares L0 and U1, respectively. Blue (hashed) and orange hatches show periods with counterclockwise and clockwise swirls.
(e) Mean Poynting flux during swirl ($\overline{S_z}|{\rm sw}$, solid) and non-swirl ($\overline{S_z}|{\rm ns}$, dashed) periods at TR1 (blue) and TR2 (orange).
(f) Upper ($R_{\rm upper}$, solid) and lower ($R_{\rm lower}$, dashed) limits of swirl contribution ratio to total Poynting flux at TR1 (blue) and TR2 (orange).
  }
  \label{fig:energy_transfer}
\end{figure*}

In this section, we analyze the transition region, i.e., the base of the corona, to quantify the Poynting flux transferred by swirls into the corona.
We apply the same swirl identification and tracking method used in the chromosphere to both sides of our simulation box, $z=3.6 \ \rm Mm$ and $z=20.4 \ \rm Mm$, hereafter referred to as TR1 and TR2, respectively.
The swirl tracking enables us to obtain the spatial and temporal coordinates of the identified swirls. 
We define the distribution function $f_{\rm sw}(x,y,t)$ of swirls as follows:

\begin{equation}
    f_{\rm sw}(x,y,t) = 
    \begin{cases}
        1 & \text{if at least one swirl exists} \\
        0 & \text{if no swirl exists}.
    \end{cases}
\end{equation}

\noindent
We then calculate the time-integrated distribution function $F_{\rm sw}(x, y)$ as follows:

\begin{equation}
    F_{\rm sw}(x, y) = \frac{\int f_{\rm sw}(x,y,t) dt}{\int f_{\rm sw}(x,y,t)dxdydt}
\end{equation}

\noindent
Figures~\ref{fig:energy_transfer}a and \ref{fig:energy_transfer}b present $F_{\rm sw}(x, y)$ at TR1 and TR2, showing distinct clusters and voids, which can be outlined with mini-squares of equal area ($4.3 \times 4.3 \ \rm Mm^2$). 
Within each mini-square, we define the period during which at least one swirl is present as the ``swirl period''. 
Furthermore, in the same region, we calculate the time evolution of the $xy$-averaged vertical Poynting flux.
Figures~\ref{fig:energy_transfer}c and \ref{fig:energy_transfer}d illustrate examples of these evolutions at TR1 and TR2, along with the corresponding swirl periods (marked by blue and orange hatches).
The results for all the mini-squares are shown in Extended Data Figures~\ref{fig:poynting_flux_tr1} and \ref{fig:poynting_flux_tr2}.
These results indicate that the Poynting flux tends to increase during swirl periods.

We then compare the Poynting flux, averaged over the horizontal plane and time, between swirl and non-swirl periods across all mini-squares (see Appendix).
Figure~\ref{fig:energy_transfer}e summarizes results for different swirl detection thresholds ($n_{\rm max}=2$--$6$).
During swirl periods, the mean Poynting flux is $3.3$--$3.7 \times 10^5 \ \rm erg \ cm^{-2} \ s^{-1}$ at TR1 and $2.3$--$2.5 \times 10^5 \ \rm erg \ cm^{-2} \ s^{-1}$ at TR2,
whereas during non-swirl periods, it drops to $0.8$--$0.9 \times 10^5$ and $0.5 \times 10^5 \ \rm erg \ cm^{-2} \ s^{-1}$, respectively.
Thus, swirl periods yield $4$--$5$ times higher flux than non-swirl periods.
Even in non-swirl periods, granulation-driven kink-mode waves contribute to coronal energy input.
This result is consistent with our earlier single-swirl event study using self-consistent simulations \citep{Kuniyoshi_2023_ApJ}.
We also estimate upper and lower bounds on the contribution of swirls to the total coronal Poynting flux (see Appendix).
As shown in Figure~\ref{fig:energy_transfer}f, the upper limit is $47$--$53\%$ at TR1 and $42$--$48\%$ at TR2, while the lower limit is $36$--$40\%$ and $34$--$38\%$, respectively.

\subsection{Coronal heating driven by swirls}
\label{sec:swirl_coronal_heating}

\begin{figure*}[htbp]
  \centering
  \includegraphics[width=15cm]{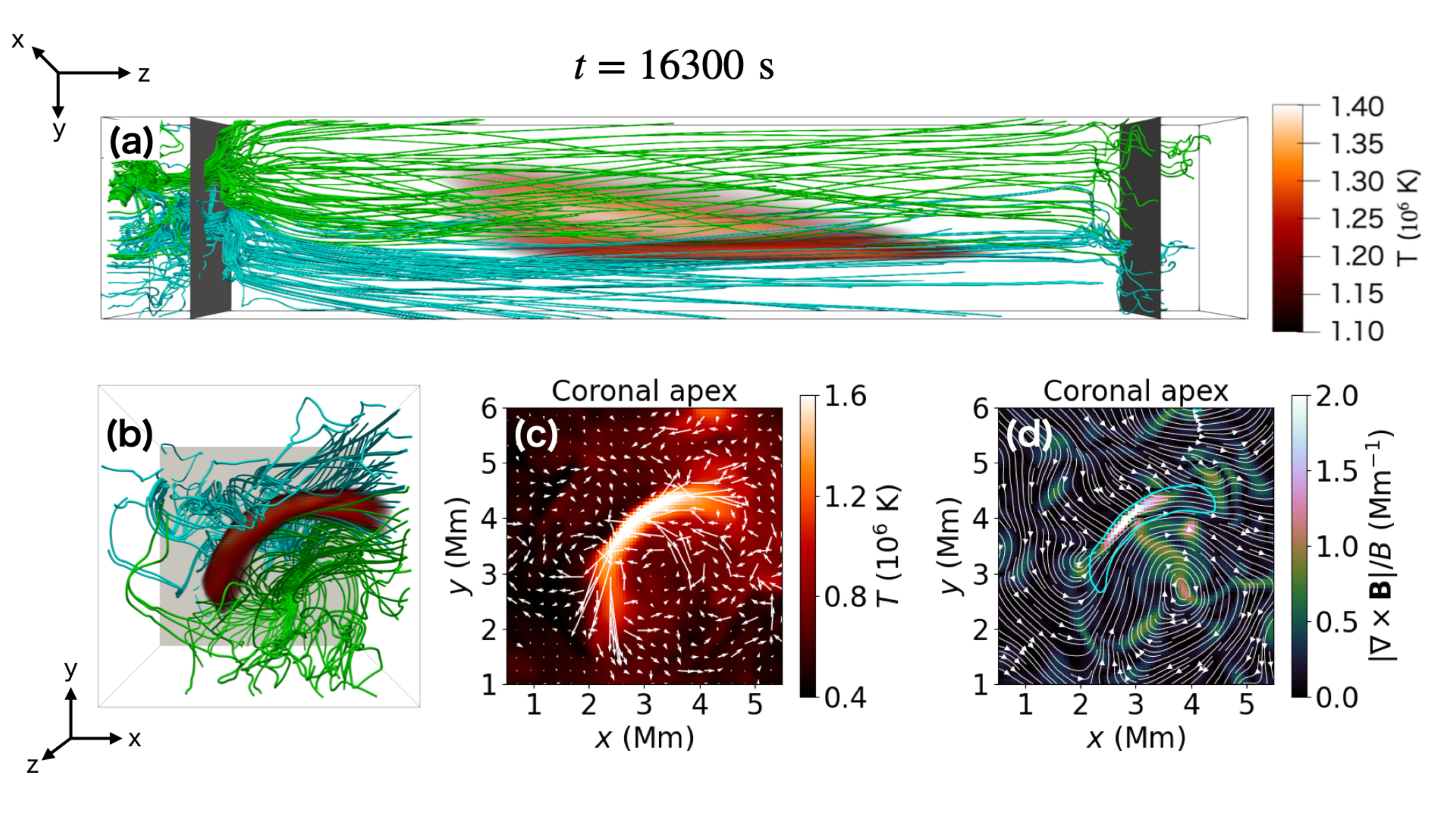} 
  \caption{
(a) and (b): Three-dimensional magnetic field lines and temperature isosurfaces associated with a coronal heating event at $t=16300 \ \rm s$.  
Green lines are twisted field lines by a swirl, and blue lines are surrounding field lines.
The field-of-view is $0.5 \ \rm Mm \le x \le 5.5 \ \rm Mm$ and $1.0 \ \rm Mm \le y \le 6.0 \ \rm Mm$, $-2.0 \ \rm Mm \le z \le 26.0 \ \rm Mm$, respectively.
Panel (a) is viewed from $x=14 \ \rm Mm$, and panel (b) from $z=24 \ \rm Mm$.
(c): Temperature with horizontal velocity $(v_x, v_y)$ at $z=12 \ \rm Mm$.
(d): $|\nabla \times \boldsymbol{B}| / B$ with horizontal magnetic field $(B_x, B_y)$, with cyan contours showing the regions of $T \ge 1.2 \ \rm MK$.
The associated animation shows the temporal evolution over a period from $t=16250 \ \rm s$ to $t=16340 \ \rm s$.
  }
  \label{fig:swirl_heating_single}
\end{figure*}

We define coronal heating events as regions where temperature exceeds $1.2 \ \rm MK$, twice the mean coronal value (see Appendix for their evolution at the coronal apex).
These events occur locally and impulsively, forming elongated structures along the $z$-direction due to thermal conduction along magnetic field lines.
As detailed below, more than $80\%$ of the coronal heating events are related to swirl events, occurring within or at the boundaries of magnetic field lines twisted by the swirls, with the location of the former exemplified in Figures~\ref{fig:swirl_heating_single}a and \ref{fig:swirl_heating_single}b.
The associated horizontal velocity fields $(v_x, v_y)$ show inflows and bi-directional outflows are generated (Figure~\ref{fig:swirl_heating_single}c).
Furthermore, strong current sheets oriented in the $z$-direction are formed either within or at the boundaries of twisted magnetic field lines (Figure~\ref{fig:swirl_heating_single}d), where the magnetic field is locally weaker than the surroundings, leading to greater values of $|\nabla \times \boldsymbol{B}|/B$.
Although indirect, these features strongly suggest that magnetic reconnection is the primary driver of these heating events.

\begin{figure*}[htbp]
  \centering
  \includegraphics[width=15cm]{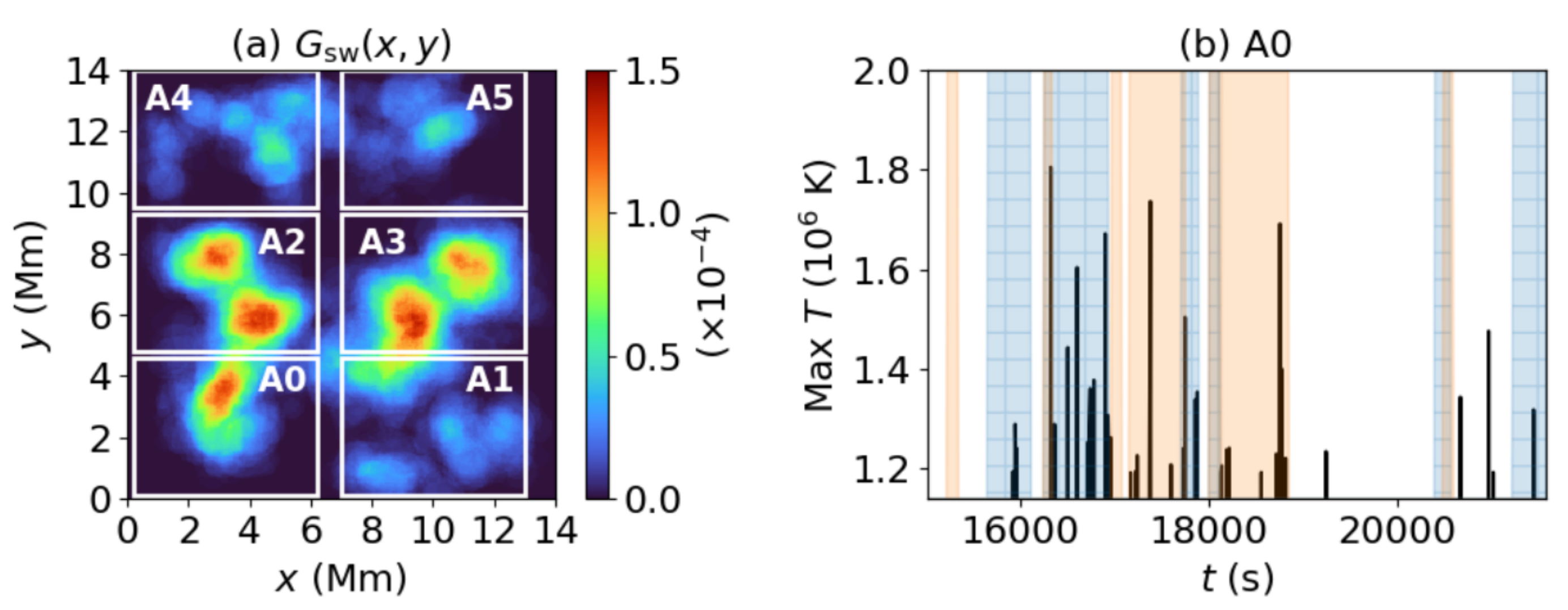} 
  \caption{
  (a) Time-integrated swirl distribution $G_{\rm sw}$ in the corona ($3.6 \ \rm Mm \le z \le 20.4 \ \rm Mm$) for $n_{\rm min}=2$, $n_{\rm max}=6$.
(b) Maximum temperatures of swirl-driven heating events within mini-square A0 in panel (a), shown at event midpoints. 
Swirl periods are marked with blue (counterclockwise) and orange (clockwise) hatches.
  }
  \label{fig:swirl_heating_group}
\end{figure*}

To investigate the correlation between the coronal heating events and swirls injected into the corona from its bases (TR1 and TR2), we define the spatiotemporal distribution function of swirls propagating into the coronal volume from its bases, $g_{\rm sw}(x,y,t)$, as follows:

\begin{equation}
    g_{\rm sw}(x,y,t) = 
    \begin{cases}
        1 & \text{if at least one swirl exists on TR1 or TR2} \\
        0 & \text{if no swirl exists on either TR1 or TR2}.
    \end{cases}
\end{equation}

\noindent
Given the Alfv{\'e}n speed in the corona, $v_{\rm A} \approx 1000 \ \rm km \ s^{-1}$, the crossing time for swirls to travel from one side of the transition region to the other as torsional Alfv{\'e}n waves is less than $20 \ \rm s$.
This timescale is effectively negligible, as it is comparable to the snapshot inverval of $10 \ \rm s$ employed in our analysis.
Thus, the time-integrated distribution function of swirls entering the corona from TR1 and TR2, denoted as $G_{\rm sw}(x, y)$, can be expressed as follows:

\begin{equation}
    G_{\rm sw}(x, y) = \frac{\int g_{\rm sw}(x,y,t) dt}{\int g_{\rm sw}(x,y,t)dxdydt} 
\end{equation}

\noindent
In Figure~\ref{fig:swirl_heating_group}a, we plot $G_{\rm sw} (x, y)$, revealing distinct clusters and voids of swirls, which are enclosed by mini-squares of equal area ($6 \times 4.5 \ \rm Mm^2$).
Because of its definition, it approximately corresponds to the superposition of $F_{\rm sw}(x, y)$ at TR1 and TR2.

The mini-squares in Figure~\ref{fig:swirl_heating_group}a represent projections of mini-boxes extending from TR1 to TR2, each with a volume of $6 \times 4.5 \times (20.4{-}3.6) \ \rm Mm^3$.
To determine whether coronal heating events occur during swirl periods, we apply the DBSCAN method \citep{Ester_1996_kddm, Einaudi_2021_ApJ} to identify 323 events in total, and track them three-dimensionally within each mini-box.
Consequently, we find that over $80\%$ of these events occur during swirl periods.
Figure~\ref{fig:swirl_heating_group}b illustrates the result for one mini-box, showing the maximum temperature within each event's isosurface at the temporal midpoint of the event, along with the corresponding swirl periods.
Similar results for other mini-boxes (see Figure~\ref{fig:hte_ssk_corona} in Appendix) consistently show that coronal heating events cluster during swirl periods, strongly indicating that swirls are their primary driver.

\begin{figure}[htbp]
  \centering
  \includegraphics[width=8.5cm]{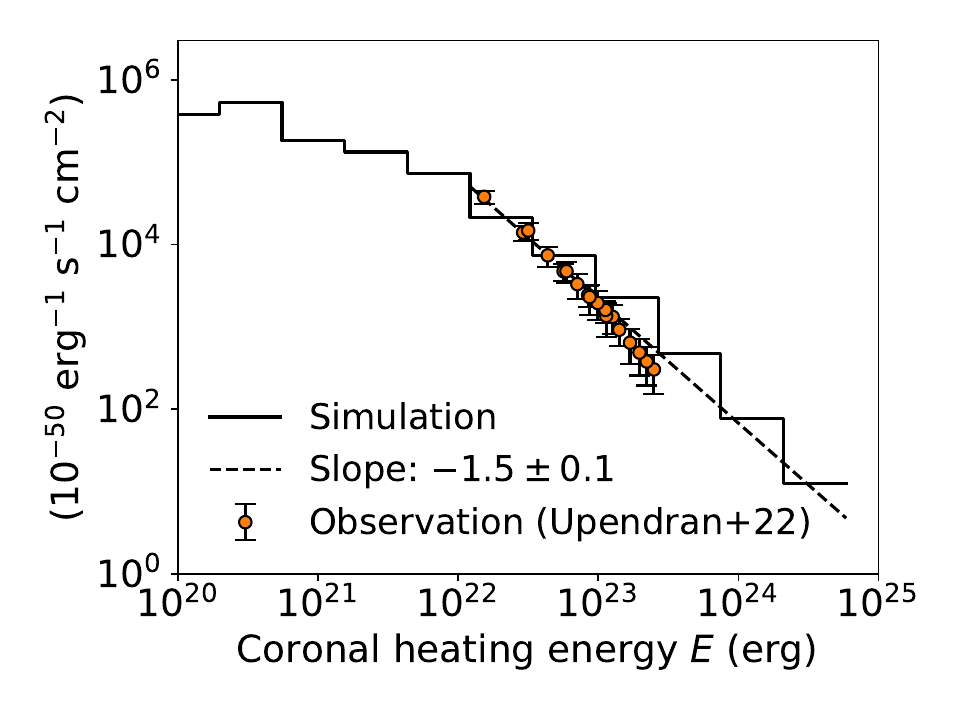} 
  \caption{Solid line: the occurrence frequency distribution of swirl-driven coronal heating energy ($dN/dE$), normalized by simulation time ($P_{\rm tot}=6600 \ \rm s$) and area ($14 \times 24 \ \rm Mm^2$).
Dashed line: fitted power-law.
Orange dots with error bars: estimated heating energies from quiet Sun coronal observation \citep{Upendran_2022_ApJ}.}
  \label{fig:power_law}
\end{figure}

To quantify how much energy is dissipated by an individual swirl-driven coronal heating event, we integrate the dissipation rate ($Q_{\rm a}$) over the event's temperature isosurface in both space and time, as expressed by:

\begin{equation}
    E = \int_{\rm event} Q_{\rm a} dx dy dz,
\end{equation}

\noindent
where $E$ corresponds to the heating energy by an individual event, and ``event'' denotes all the voxels in space and time that belong to the event.
We derive the occurrence frequency of the swirl-driven coronal heating events as a function of $E$, denoted as $dN/dE$ where $N$ is the number of the events.
Figure~\ref{fig:power_law} illustrates the distribution of $dN/dE$ normalized by $P_{\rm tot} L_{\perp} L_z$, where $P_{\rm tot} = 6600 \ \rm s$ is the simulation duration, $L_{\perp} = 14 \ \rm Mm$ and $L_z = 24 \ \rm Mm$ are the horizontal and vertical size of the simulated atmosphere, respectively.
This normalization corresponds to observations along the line of sight in the horizontal direction (either the $x$- or $y$-direction).
Within the energy range of $10^{22}$--$10^{25} \ \rm erg$, $dN/dE$ exhibits a power-law-like behavior, i.e.,

\begin{equation}
    dN/dE \propto E^{- \alpha},
\end{equation}

\noindent
where $\alpha$ is a power-law index.
Fitting the curve yields $\alpha= 1.5 \pm 0.1$ for histograms with bin numbers ranging from $10$ to $25$.
In contrast, the slope of $dN/dE$ for \added{$E < 10^{22} \ \rm erg$} is flatter. 
However, this does not indicate a low number of coronal heating events within this range; rather, our identification method is unable to effectively isolate such small-scale events, as they do not heat the corona beyond our detection threshold ($T \geq 1.2 \ \rm MK$).
Figure~\ref{fig:power_law} also presents previous observational results of coronal heating events in the quiet Sun corona within the energy range of $10^{22}$--$10^{25} \ \rm erg$ \citep{Upendran_2022_ApJ}, demonstrating a close match between our simulated distribution and the observed ones.
This result indicates that swirls are promising candidates for the drivers of coronal heating events, at least within the nanoflare energy regime.

\section{Summary and discussions}\label{sec:discussion}

Our numerical analysis statistically demonstrates the critical role of swirls in heating the quiet Sun corona. First, swirl-driven torsional Alfv{\'e}nic waves are responsible for transporting approximately half of the total magnetic energy into the corona (Figure~\ref{fig:energy_transfer}f). 
Second, swirl-driven coronal heating reproduces the statistical properties of nanoflare-like heating events, i.e., their energy frequency distribution (Figure~\ref{fig:power_law}), suggesting that swirls act as their drivers.
Additionally, swirl-driven heating occurs via magnetic reconnection within current sheets formed at misalignments of twisted magnetic field lines (Figure~\ref{fig:swirl_heating_single}c--f), \added{a process similar to that found in the previous simulation \citep{Galsgaard_1996_JGR}.}

\added{It should be noted that our results differ from several previous self-consistent coronal heating simulations that include photospheric convective motions, proposing that granular motions are the main contributor for magnetic energy supply into the corona \citep{Gudiksen_2005_ApJ, Rempel_2017_ApJ}. This discrepancy likely stems from their limited spatial resolution (several hundred kilometers) which cannot resolve intergranular lanes, leading to numerical dissipation of photospheric swirls before they can fully develop. As a result, the swirl contribution may be underestimated.
By contrast, we consider that our results align with the self-consistent simulation by \citet{Hansteen_2015_ApJ}, which employs a similar numerical resolution ($65\ \mathrm{km}$). Although they do not explicitly identify the roles of specific photospheric motions (i.e., granular or swirling motions), we suggest that swirls significantly contribute to magnetic energy supply. 
Indeed, in another simulation—with the same code, box size, spatial resolution, and similar unsigned photospheric magnetic field strength, though focused on an open flux region—swirls are clearly generated and contribute to the coronal energy supply \citep{Silva_2024_ApJ}.}

Our study presents a direct comparison between simulated and observed chromospheric swirl properties, showing good agreement with previous observations \citep{Dakanalis_2022_AA}.
However, our results generally yield smaller values (see Section~\ref{sec:swirl_properties}), likely due to differences in swirl identification methods.
\citet{Dakanalis_2022_AA} relied on morphological features of swirls, as extracting in-plane velocity fields from observational data is challenging.
To resolve these discrepancies, it is crucial to apply the observational identification method to the synthesized chromosphere from our simulation.

The simulated swirls are concentrated in several distinct locations (Figure~\ref{fig:energy_transfer}a, b). 
This result may be attributed to the organization of the photospheric magnetic field by mesogranulation, an intermediate convective scale between granulation and supergranulation that has been identified both observationally and numerically \citep{November_1981_ApJ, Amari_2015_Natur}.
The distances between the simulated swirl clusters are consistent with the typical diameter of a mesogranular cell \citep[$\sim 5$ Mm;][]{November_1981_ApJ}.
In contrast, the typical lifetime of mesogranulation is approximately $2 \ \rm hr $ \citep{November_1981_ApJ}, which exceeds our simulation duration.
This may explain why the Poynting flux is higher at TR1 than at TR2 (Figure~\ref{fig:energy_transfer}f).

\added{Our numerical model has several limitations.
First, the setup represents the quiet Sun network region, which is primarily composed of unipolar magnetic fields and lacks internetwork regions with low-lying bipolar fields extending up to chromospheric or transition region heights \citep{BellotRubio_2019_LRSP}.
Such bipolar fields may also contribute to the magnetic energy supply to the quiet Sun corona, which should be addressed by a more comprehensive model that includes both network and internetwork regions.
However, we believe the absence of the internetwork region does not affect our main conclusion, as previous simulations show the network dominates magnetic energy supply to the bulk of the corona, while the internetwork contributes only up to the transition region \citep{Amari_2015_Natur}.
Second, our numerical resolution $60 \ \rm km$ may not sufficient to reproduce the actual photospheric convection. Higher resolution allows for a more accurate representation of velocity fields within intergranular lanes. This, in turn, helps prevent swirls from being prematurely dissipated before they can fully develop. As a result, a greater number of swirls may be able to transfer magnetic energy into the corona.}


\begin{acknowledgments}

The authors would like to thank Haruhisa Iijima for providing the RAMENS code.
Numerical simulations were carried out using the Cray XC50 system at the Center for Computational Astrophysics (CfCA), National Astronomical Observatory of Japan, and the A-KDK computer system at the Research Institute for Sustainable Humanosphere, Kyoto University.
H.K. gratefully acknowledges support from the JSPS (Japan Society for the Promotion of Science) Overseas Research Fellowship.
S.I. is supported by JSPS KAKENHI Grant Numbers 25K01052, 25K00976, and 24K00688.
T.Y. is supported by JSPS KAKENHI Grant Number JP21H04492.
This research is also supported by the Joint Research Program of the National Institutes of Natural Sciences (NINS), Grant Number OML032402.
This research is also sponsored by the DynaSun project and has thus received funding under the Horizon Europe programme of the European Union under grant agreement (no. 101131534). 
Views and opinions expressed are however those of the authors only and do not necessarily reflect those of the European Union and therefore the European Union cannot be held responsible for them. 
This work is also supported by the Engineering and Physical Sciences Research Council (EP/Y037464/1) under the Horizon Europe Guarantee.

\end{acknowledgments}

\appendix

\section{Estimation of heating rate}

We do not include explicit viscosity or resistivity in the governing equations; therefore, all heating processes arise from numerical dissipation.
As the finite volume method is employed, the internal energy $e_{\rm int}$ is computed after updating all the conservative variables, including mass density $\rho$, momentum $\rho \boldsymbol{v}$, magnetic field $\boldsymbol{B}$ and total energy $e=e_{\rm int} + \rho \boldsymbol{v}^2/2 + \boldsymbol{B}^2/8\pi$.
Therefore, the internal energy equation (identified by the subscript ``tot'') is obtained as follows:

\begin{equation}\label{eq:eint_tot}
    \left( \frac{\partial e_{\rm int}}{\partial t} \right)_{\rm tot}
    = \frac{\partial e}{\partial t} 
    - \frac{\partial}{\partial t} \left( \frac{(\rho \boldsymbol{v}^2)}{2 \rho} + \frac{\boldsymbol{B}^2}{8 \pi} \right).
\end{equation}

\noindent
On the other hand, the direct expression of the internal energy equation (identified by the subscript ``ad'') is written as:

\begin{equation}\label{eq:eint_ad}
    \left( \frac{\partial e_{\rm int}}{\partial t} \right)_{\rm ad} = -\nabla \cdot (e_{\rm int} \boldsymbol{v})
    -p \nabla \cdot \boldsymbol{v} + Q_{\rm cnd} + Q_{\rm rad},
\end{equation}

\noindent
where $p$ is gas pressure, $Q_{\rm cnd}$ is the heating rate by Spitzer-H{\"a}rm thermal conduction along magnetic field lines \citep{Spitzer_1953_PhRv}, and $Q_{\rm rad}$ denotes the heating rate due to radiation.
The deviation of $Q_{\rm cnd}$ and $Q_{\rm rad}$ is described in \citet{Iijima_2016_PhDT}.
Analytically, Equations~\eqref{eq:eint_tot} and \eqref{eq:eint_ad} should be identical. 
However, they are not because the numerical diffusion associated with the upwind scheme in the momentum and induction equations results in dissipative heating.
Due to numerical dissipation, the following inequality should be satisfied:

\begin{equation}\label{eq:inequality_eint_tot_eint_ad}
   \left( \frac{\partial e_{\rm int}}{\partial t} \right)_{\rm tot} \ge \left( \frac{\partial e_{\rm int}}{\partial t} \right)_{\rm ad}.
\end{equation}

\noindent
Therefore, following \citet{Matsumoto_2014_MNRAS}, we evaluate the numerical dissipation term $Q_{\rm a}$ as:

\begin{equation}
    Q_{\rm a} = \left( \frac{\partial e_{\rm int}}{\partial t} \right)_{\rm tot} - \left( \frac{\partial e_{\rm int}}{\partial t} \right)_{\rm ad}.
\end{equation}

\noindent
It is important to note that $Q_{\rm a}$ is evaluated at each individual time step during the computation.
In the practical case, Equation~\eqref{eq:inequality_eint_tot_eint_ad} is not strictly satisfied because of the discretization errors, leading to negative $Q_{\rm a}$.
However, this effect is negligible and does not affect our conclusions.

\section{Poynting flux carried by swirls}\label{subsec:poynting_flux_swirls}

We calculate the time evolution of the $xy$-averaged vertical Poynting flux at TR1 and TR2 within each mini-square (Figures~\ref{fig:energy_transfer}a and \ref{fig:energy_transfer}b) as follows:

\begin{equation}
    \langle S_z \rangle_{m} = \frac{1}{a_{\rm sq}} \int_{m} S_z dx dy,
\end{equation}

\noindent
where ``$m$'' refers to the areas outlined by each mini-square (L0--L7 in TR1 and U0--U7 in TR2) and $a_{\rm sq}$ is the area of each mini-square, given by $4.3 \times 4.3 \ \rm Mm^2$.
We then calculate the mean values of the Poynting flux during the swirl periods over the mini-squares ($\overline{S_z} |_{\rm{sw}}$) as follows:

\begin{align}
    & \overline{S_z} |_{\rm{sw}, \it{m}} = \frac{\sum_k \int_k \langle S_z \rangle_{m} dt}{\tau_{\rm{sw}, \it{m}}}, \quad \tau_{\rm{sw}, \it{m}} = \sum_{k} \int_k dt, \\
    & \overline{S_z} |_{\rm sw} = \frac{\sum_{m} \tau_{\rm{sw}, \it{m}} \overline{S_z} |_{\rm{sw}, \it{m}}}{\sum_{m} \tau_{\rm{sw}, \it{m}}},
\end{align}

\noindent
where $k$ represents the individual swirl periods.
In addition, we compute the mean values of the Poynting flux during the non-swirl periods over the mini-squares ($\overline{S_z} |_{\rm{ns}}$) as follows:

\begin{align}
    & \overline{S_z} |_{\rm{ns}, \it{m}} = \frac{\sum_l \int_l \langle S_z \rangle_{m} dt}{\tau_{\rm{ns}, \it{m}}}, \quad \tau_{\rm{ns}, \it{m}} = \sum_{l} \int_l dt, \\
    & \overline{S_z} |_{\rm ns} = \frac{\sum_{m'} \tau_{\rm{ns}, \it{m'}} \overline{S_z} |_{\rm{ns}, \it{m'}}}{\sum_{m'} \tau_{\rm{ns}, \it{m'}}},
\end{align}

\noindent
where $l$ represents the individual non-swirl periods, and ``$m'$'' corresponds to L1, L6, L7 for TR1 and U0, U2 for TR2.
These are the areas where the non-swirl periods account for at least $80\%$, irrespective of the swirl detection threshold.
We define the value $\overline{S_z} |_{\rm{ns}}$ for the case of $n_{\rm min} = n_{\rm max} = 2$ as representing the basal flux ($\overline{S_z}|_{\rm{base}}$) transported by the non-swirl-driven mechanism (i.e., kink-mode waves).

Furthermore, we estimate the contribution rate of swirls to the total amount of Poynting flux into the corona.
We set the upper limit ($R_{\rm upper}$) based on the Poynting flux during the swirl periods, and the lower limit ($R_{\rm lower}$) as the Poynting flux during those periods minus the basal flux as follows:

\begin{align}
    & R_{\rm upper} = \frac{\sum_{m} a_{\rm sq} \tau_{\rm{sw}, \it{m}} \overline{S_z}|_{\rm{sw}, \it{m}}}{\int S_z dx dy dt}, \\
    & R_{\rm lower} = \frac{\sum_{m} a_{\rm sq} \tau_{\rm{sw}, \it{m}} \left( \overline{S_z}|_{\rm{sw}, \it{m}} - \overline{S_z}|_{\rm{base}} \right)}{\int S_z dx dy dt}.
\end{align}

\section{Additional figures}\label{subsec:additional_figures}

\begin{figure}[htbp]
  \centering
  \includegraphics[width=13cm]{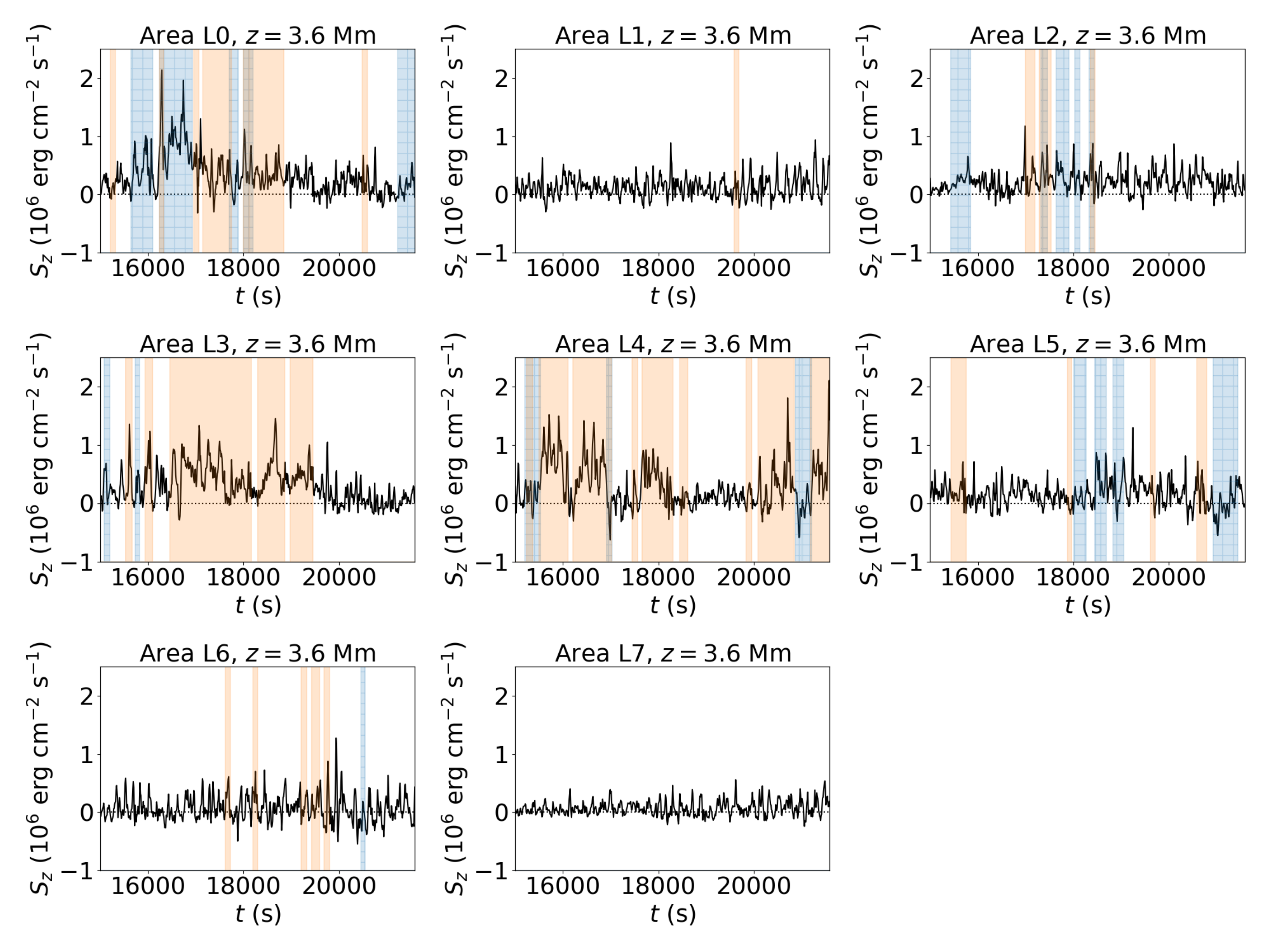} 
  \caption{
  Time evolution of $xy$-averaged Poynting flux within the mini-squares (L0--L7) outlined in the swirl distribution map at TR1 (Figure~\ref{fig:energy_transfer}a) in the case of $n_{\rm max}=6$. 
  Blue (hashed) and orange hatches indicate periods when swirls are present within the box, with counterclockwise and clockwise rotation angles, respectively.
  }
  \label{fig:poynting_flux_tr1}
\end{figure}

\begin{figure}[htbp]
  \centering
  \includegraphics[width =13cm]{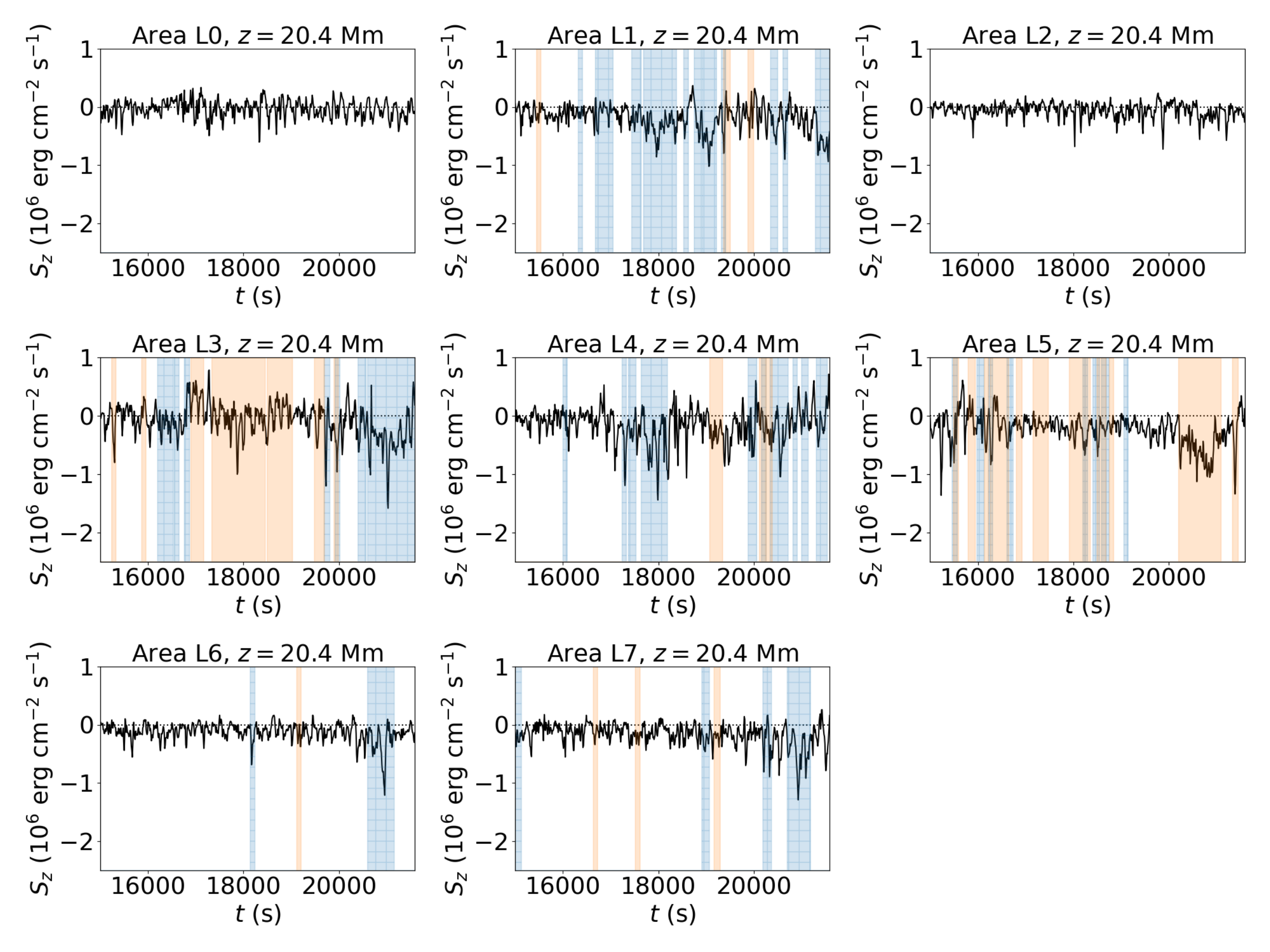}
  \caption{
  Same as Figure~\ref{fig:poynting_flux_tr1} but showing $xy$-averaged Poynting flux within the mini-squares (U0--U7) at TR2 (Figure~\ref{fig:energy_transfer}b).
  }
\label{fig:poynting_flux_tr2}
\end{figure}

\begin{figure}[htbp]
  \centering
  \includegraphics[width =8.5cm]{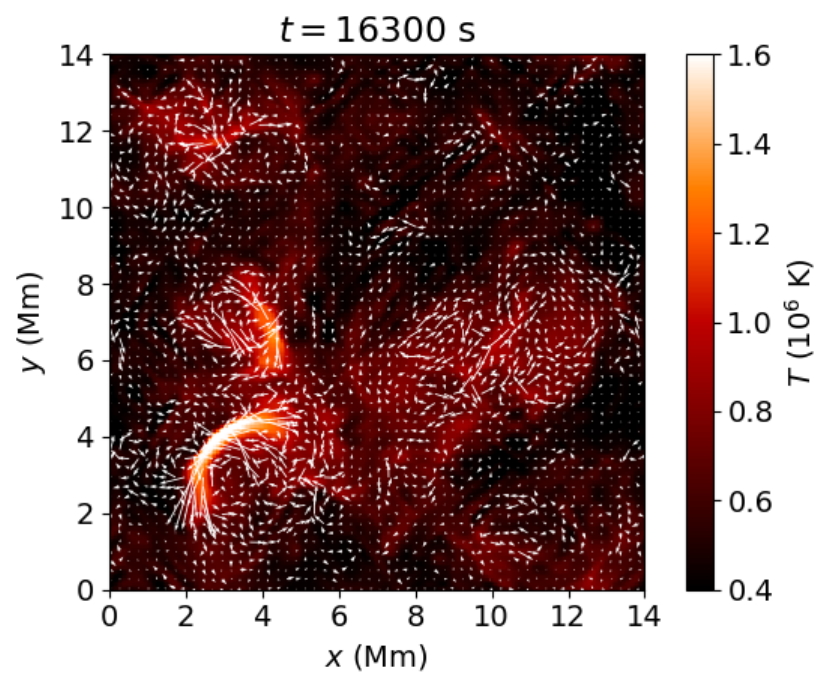}
  \caption{
  A snapshot of temperature $T$ at the coronal apex, with the horizontal velocity field $(v_x,v_y)$ shown as arrows.
  The associated animation shows the temporal evolution over a period from $t=15000 \ \rm s$ to $t=21600 \ \rm s$.
  }
\label{fig:te_apex}
\end{figure}

\begin{figure}[htbp]
  \centering
  \includegraphics[width =15cm]{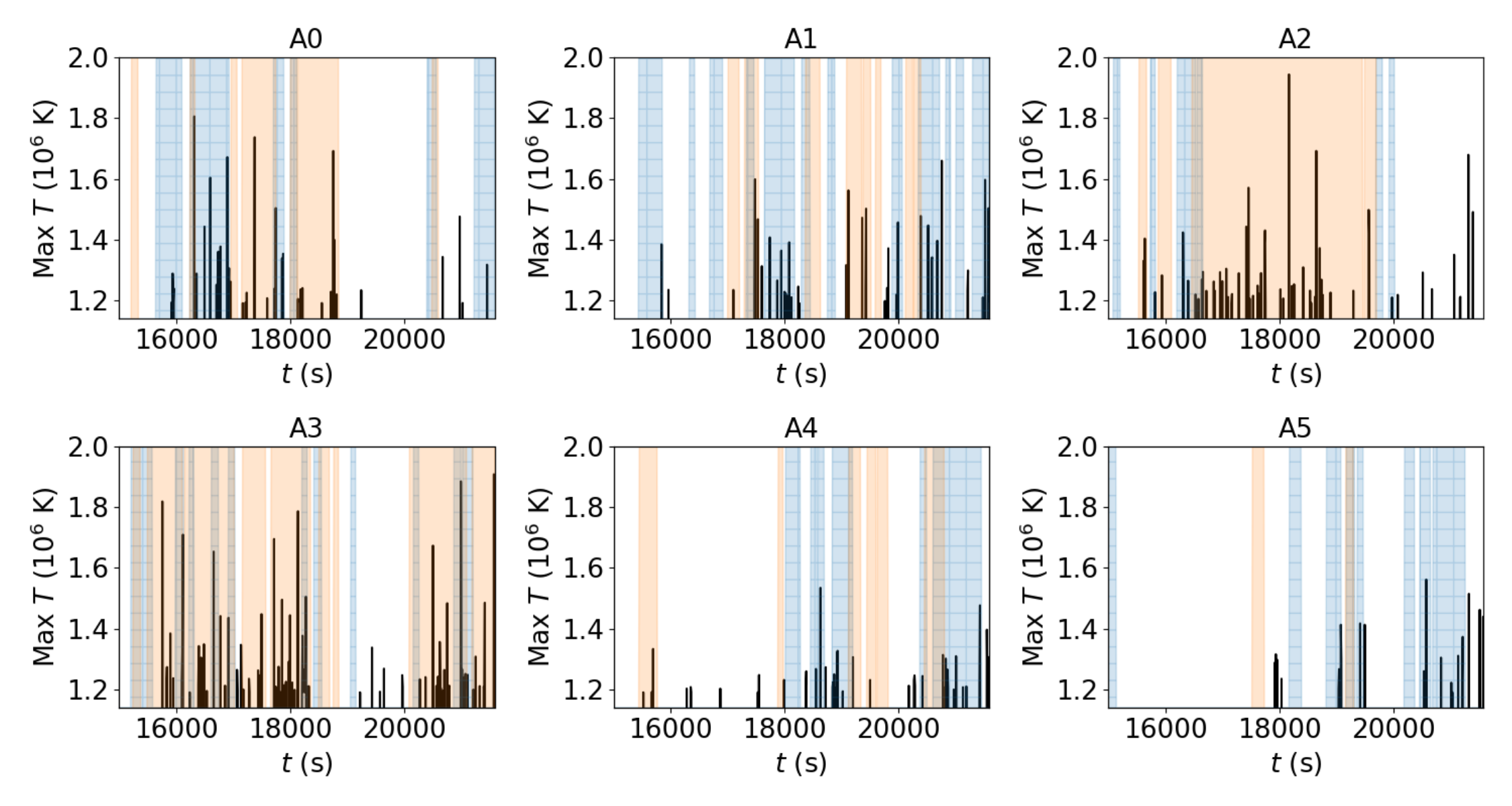}
  \caption{
  The maximum temperatures of the coronal heating events at the midpoints between the start and end times for the case with $n_{\rm max}=6$ are plotted within each mini-square (A0--A5) shown in Figure~\ref{fig:swirl_heating_group}a.
  Simultaneously, the swirl periods are shown as blue (hashed) and orange hatches, indicating counterclockwise and clockwise rotation directions, respectively.
  }
\label{fig:hte_ssk_corona}
\end{figure}

Figures~\ref{fig:poynting_flux_tr1} and \ref{fig:poynting_flux_tr2} show the time evolution of $xy$-averaged Poynting flux at TR1 and TR2 within the mini-squares (L0--L7 and U0--U7) defined in Figures~\ref{fig:energy_transfer}a and \ref{fig:energy_transfer}b, respectively.
Figure~\ref{fig:te_apex} presents a snapshot of the cross-sectional temperature distribution at the coronal apex, overlaid with horizontal velocity vectors $(v_x, v_y)$.
Figure~\ref{fig:hte_ssk_corona} displays the maximum temperatures of coronal heating events at their temporal midpoints within each mini-square (A0–A5) shown in Figure~\ref{fig:swirl_heating_group}a.

\clearpage



\begin{thebibliography}{}
\expandafter\ifx\csname natexlab\endcsname\relax\def\natexlab#1{#1}\fi
\providecommand{\url}[1]{\href{#1}{#1}}
\providecommand{\dodoi}[1]{doi:~\href{http://doi.org/#1}{\nolinkurl{#1}}}
\providecommand{\doeprint}[1]{\href{http://ascl.net/#1}{\nolinkurl{http://ascl.net/#1}}}
\providecommand{\doarXiv}[1]{\href{https://arxiv.org/abs/#1}{\nolinkurl{https://arxiv.org/abs/#1}}}

\bibitem[{T. {Amari} {et~al.}(2015){Amari}, {Luciani}, \& {Aly}}]{Amari_2015_Natur}
{Amari}, T., {Luciani}, J.-F., \& {Aly}, J.-J. 2015, \bibinfo{title}{{Small-scale dynamo magnetism as the driver for heating the solar atmosphere},} \nat, 522, 188, \dodoi{10.1038/nature14478}

\bibitem[{L. {Bellot Rubio} \& D. {Orozco Su{\'a}rez}(2019){Bellot Rubio} \& {Orozco Su{\'a}rez}}]{BellotRubio_2019_LRSP}
{Bellot Rubio}, L., \& {Orozco Su{\'a}rez}, D. 2019, \bibinfo{title}{{Quiet Sun magnetic fields: an observational view},} Living Reviews in Solar Physics, 16, 1, \dodoi{10.1007/s41116-018-0017-1}

\bibitem[{T.~E. {Berger} {et~al.}(1998){Berger}, {L{\"o}fdahl}, {Shine}, \& {Title}}]{Berger_1998_ApJ}
{Berger}, T.~E., {L{\"o}fdahl}, M.~G., {Shine}, R.~S., \& {Title}, A.~M. 1998, \bibinfo{title}{{Measurements of Solar Magnetic Element Motion from High-Resolution Filtergrams},} \apj, 495, 973, \dodoi{10.1086/305309}

\bibitem[{J.~A. {Bonet} {et~al.}(2008){Bonet}, {M{\'a}rquez}, {S{\'a}nchez Almeida}, {Cabello}, \& {Domingo}}]{Bonet_2008_ApJ}
{Bonet}, J.~A., {M{\'a}rquez}, I., {S{\'a}nchez Almeida}, J., {Cabello}, I., \& {Domingo}, V. 2008, \bibinfo{title}{{Convectively Driven Vortex Flows in the Sun},} \apjl, 687, L131, \dodoi{10.1086/593329}

\bibitem[{C. {Breu} {et~al.}(2022){Breu}, {Peter}, {Cameron}, {Solanki}, {Przybylski}, {Rempel}, \& {Chitta}}]{Breu_2022_AA}
{Breu}, C., {Peter}, H., {Cameron}, R., {et~al.} 2022, \bibinfo{title}{{A solar coronal loop in a box: Energy generation and heating},} \aap, 658, A45, \dodoi{10.1051/0004-6361/202141451}

\bibitem[{P.~S. {Cally} \& M. {Goossens}(2008){Cally} \& {Goossens}}]{Cally_2008_SoPh}
{Cally}, P.~S., \& {Goossens}, M. 2008, \bibinfo{title}{{Three-Dimensional MHD Wave Propagation and Conversion to Alfv{\'e}n Waves near the Solar Surface. I. Direct Numerical Solution},} \solphys, 251, 251, \dodoi{10.1007/s11207-007-9086-3}

\bibitem[{J.~R. {Canivete Cuissa} \& O. {Steiner}(2020){Canivete Cuissa} \& {Steiner}}]{CaniveteCuissa_2020_AA}
{Canivete Cuissa}, J.~R., \& {Steiner}, O. 2020, \bibinfo{title}{{Vortices evolution in the solar atmosphere. A dynamical equation for the swirling strength},} \aap, 639, A118, \dodoi{10.1051/0004-6361/202038060}

\bibitem[{L.~P. {Chitta} {et~al.}(2021){Chitta}, {Peter}, \& {Young}}]{Chitta_2021_AA}
{Chitta}, L.~P., {Peter}, H., \& {Young}, P.~R. 2021, \bibinfo{title}{{Extreme-ultraviolet bursts and nanoflares in the quiet-Sun transition region and corona},} \aap, 647, A159, \dodoi{10.1051/0004-6361/202039969}

\bibitem[{M.~S. {Chong} {et~al.}(1990){Chong}, {Perry}, \& {Cantwell}}]{Chong_1990_PhFlA}
{Chong}, M.~S., {Perry}, A.~E., \& {Cantwell}, B.~J. 1990, \bibinfo{title}{{A general classification of three-dimensional flow fields},} Physics of Fluids A, 2, 765, \dodoi{10.1063/1.857730}

\bibitem[{J. {Christensen-Dalsgaard} {et~al.}(1996){Christensen-Dalsgaard}, {Dappen}, {Ajukov}, {Anderson}, {Antia}, {Basu}, {Baturin}, {Berthomieu}, {Chaboyer}, {Chitre}, {Cox}, {Demarque}, {Donatowicz}, {Dziembowski}, {Gabriel}, {Gough}, {Guenther}, {Guzik}, {Harvey}, {Hill}, {Houdek}, {Iglesias}, {Kosovichev}, {Leibacher}, {Morel}, {Proffitt}, {Provost}, {Reiter}, {Rhodes}, {Rogers}, {Roxburgh}, {Thompson}, \& {Ulrich}}]{ChristensenDalsgaard_1996_Sci}
{Christensen-Dalsgaard}, J., {Dappen}, W., {Ajukov}, S.~V., {et~al.} 1996, \bibinfo{title}{{The Current State of Solar Modeling},} Science, 272, 1286, \dodoi{10.1126/science.272.5266.1286}

\bibitem[{S.~R. {Cranmer}(2018){Cranmer}}]{Cranmer_2018_ApJ}
{Cranmer}, S.~R. 2018, \bibinfo{title}{{Low-frequency Alfv{\'e}n Waves Produced by Magnetic Reconnection in the Sun{\textquoteright}s Magnetic Carpet},} \apj, 862, 6, \dodoi{10.3847/1538-4357/aac953}

\bibitem[{S.~R. {Cranmer} \& A.~A. {van Ballegooijen}(2005){Cranmer} \& {van Ballegooijen}}]{Cranmer_2005_ApJS}
{Cranmer}, S.~R., \& {van Ballegooijen}, A.~A. 2005, \bibinfo{title}{{On the Generation, Propagation, and Reflection of Alfv{\'e}n Waves from the Solar Photosphere to the Distant Heliosphere},} \apjs, 156, 265, \dodoi{10.1086/426507}

\bibitem[{I. {Dakanalis} {et~al.}(2022){Dakanalis}, {Tsiropoula}, {Tziotziou}, \& {Kontogiannis}}]{Dakanalis_2022_AA}
{Dakanalis}, I., {Tsiropoula}, G., {Tziotziou}, K., \& {Kontogiannis}, I. 2022, \bibinfo{title}{{Chromospheric swirls. I. Automated detection in H{\ensuremath{\alpha}} observations and their statistical properties},} \aap, 663, A94, \dodoi{10.1051/0004-6361/202243236}

\bibitem[{K.~P. {Dere} {et~al.}(1997){Dere}, {Landi}, {Mason}, {Monsignori Fossi}, \& {Young}}]{Dere_1997_AAS}
{Dere}, K.~P., {Landi}, E., {Mason}, H.~E., {Monsignori Fossi}, B.~C., \& {Young}, P.~R. 1997, \bibinfo{title}{{CHIANTI - an atomic database for emission lines},} \aaps, 125, 149, \dodoi{10.1051/aas:1997368}

\bibitem[{G. {Einaudi} {et~al.}(2021){Einaudi}, {Dahlburg}, {Ugarte-Urra}, {Reep}, {Rappazzo}, \& {Velli}}]{Einaudi_2021_ApJ}
{Einaudi}, G., {Dahlburg}, R.~B., {Ugarte-Urra}, I., {et~al.} 2021, \bibinfo{title}{{Energetics and 3D Structure of Elementary Events in Solar Coronal Heating},} \apj, 910, 84, \dodoi{10.3847/1538-4357/abe464}

\bibitem[{M. {Ester} {et~al.}(1996){Ester}, {Kriegel}, {Sander}, \& {Xu}}]{Ester_1996_kddm}
{Ester}, M., {Kriegel}, H.-P., {Sander}, J., \& {Xu}, X. 1996, \bibinfo{title}{{A Density-Based Algorithm for Discovering Clusters in Large Spatial Databases with Noise},} in Second International Conference on Knowledge Discovery and Data Mining (KDD'96). Proceedings of a conference held August 2-4, ed. D.~W. {Pfitzner} \& J.~K. {Salmon}, 226--331

\bibitem[{K. {Galsgaard} \& {\r{A}}. {Nordlund}(1996){Galsgaard} \& {Nordlund}}]{Galsgaard_1996_JGR}
{Galsgaard}, K., \& {Nordlund}, {\r{A}}. 1996, \bibinfo{title}{{Heating and activity of the solar corona 1. Boundary shearing of an initially homogeneous magnetic field},} \jgr, 101, 13445, \dodoi{10.1029/96JA00428}

\bibitem[{M.~L. {Goodman} \& P.~G. {Judge}(2012){Goodman} \& {Judge}}]{Goodman_2012_ApJ}
{Goodman}, M.~L., \& {Judge}, P.~G. 2012, \bibinfo{title}{{Radiating Current Sheets in the Solar Chromosphere},} \apj, 751, 75, \dodoi{10.1088/0004-637X/751/1/75}

\bibitem[{B.~V. {Gudiksen} \& {\r{A}}. {Nordlund}(2005){Gudiksen} \& {Nordlund}}]{Gudiksen_2005_ApJ}
{Gudiksen}, B.~V., \& {Nordlund}, {\r{A}}. 2005, \bibinfo{title}{{An Ab Initio Approach to the Solar Coronal Heating Problem},} \apj, 618, 1020, \dodoi{10.1086/426063}

\bibitem[{V. {Hansteen} {et~al.}(2015){Hansteen}, {Guerreiro}, {De Pontieu}, \& {Carlsson}}]{Hansteen_2015_ApJ}
{Hansteen}, V., {Guerreiro}, N., {De Pontieu}, B., \& {Carlsson}, M. 2015, \bibinfo{title}{{Numerical Simulations of Coronal Heating through Footpoint Braiding},} \apj, 811, 106, \dodoi{10.1088/0004-637X/811/2/106}

\bibitem[{H. {Iijima}(2016){Iijima}}]{Iijima_2016_PhDT}
{Iijima}, H. 2016, \bibinfo{title}{{Numerical studies of solar chromospheric jets},} PhD thesis, University of Tokyo, Department of Earth and Planetary Environmental Science

\bibitem[{P.~G. {Judge}(2023){Judge}}]{Judge_2023_ApJ}
{Judge}, P.~G. 2023, \bibinfo{title}{{Steadiness of Coronal Heating},} \apj, 957, 25, \dodoi{10.3847/1538-4357/acf83a}

\bibitem[{Y. {Kato} \& S. {Wedemeyer}(2017){Kato} \& {Wedemeyer}}]{Kato_2017_AA}
{Kato}, Y., \& {Wedemeyer}, S. 2017, \bibinfo{title}{{Vortex flows in the solar chromosphere. I. Automatic detection method},} \aap, 601, A135, \dodoi{10.1051/0004-6361/201630082}

\bibitem[{T. {Kawai} \& S. {Imada}(2021){Kawai} \& {Imada}}]{Kawai_2021_ApJ}
{Kawai}, T., \& {Imada}, S. 2021, \bibinfo{title}{{Energy Distribution of Small-scale Flares Derived Using a Genetic Algorithm},} \apj, 906, 2, \dodoi{10.3847/1538-4357/abc9ae}

\bibitem[{J.~A. {Klimchuk}(2006){Klimchuk}}]{Klimchuk_2006_SoPh}
{Klimchuk}, J.~A. 2006, \bibinfo{title}{{On Solving the Coronal Heating Problem},} \solphys, 234, 41, \dodoi{10.1007/s11207-006-0055-z}

\bibitem[{H. {Kuniyoshi} {et~al.}(2024){Kuniyoshi}, {Bose}, \& {Yokoyama}}]{Kuniyoshi_2024_ApJ}
{Kuniyoshi}, H., {Bose}, S., \& {Yokoyama}, T. 2024, \bibinfo{title}{{Comprehensive Synthesis of Magnetic Tornado: Cospatial Incidence of Chromospheric Swirls and Extreme-ultraviolet Brightening},} \apjl, 969, L34, \dodoi{10.3847/2041-8213/ad5a0e}

\bibitem[{H. {Kuniyoshi} {et~al.}(2023){Kuniyoshi}, {Shoda}, {Iijima}, \& {Yokoyama}}]{Kuniyoshi_2023_ApJ}
{Kuniyoshi}, H., {Shoda}, M., {Iijima}, H., \& {Yokoyama}, T. 2023, \bibinfo{title}{{Magnetic Tornado Properties: A Substantial Contribution to the Solar Coronal Heating via Efficient Energy Transfer},} \apj, 949, 8, \dodoi{10.3847/1538-4357/accbb8}

\bibitem[{E. {Landi} {et~al.}(2012){Landi}, {Del Zanna}, {Young}, {Dere}, \& {Mason}}]{Landi_2012_ApJ}
{Landi}, E., {Del Zanna}, G., {Young}, P.~R., {Dere}, K.~P., \& {Mason}, H.~E. 2012, \bibinfo{title}{{CHIANTI{\textemdash}An Atomic Database for Emission Lines. XII. Version 7 of the Database},} \apj, 744, 99, \dodoi{10.1088/0004-637X/744/2/99}

\bibitem[{J. {Liu} {et~al.}(2019){Liu}, {Nelson}, {Snow}, {Wang}, \& {Erd{\'e}lyi}}]{Liu_2019_NatCo}
{Liu}, J., {Nelson}, C.~J., {Snow}, B., {Wang}, Y., \& {Erd{\'e}lyi}, R. 2019, \bibinfo{title}{{Evidence of ubiquitous Alfv{\'e}n pulses transporting energy from the photosphere to the upper chromosphere},} Nature Communications, 10, 3504, \dodoi{10.1038/s41467-019-11495-0}

\bibitem[{T. {Matsumoto} \& T.~K. {Suzuki}(2014){Matsumoto} \& {Suzuki}}]{Matsumoto_2014_MNRAS}
{Matsumoto}, T., \& {Suzuki}, T.~K. 2014, \bibinfo{title}{{Connecting the Sun and the solar wind: the self-consistent transition of heating mechanisms},} \mnras, 440, 971, \dodoi{10.1093/mnras/stu310}

\bibitem[{R.~J. {Morton} {et~al.}(2025){Morton}, {Molnar}, {Cranmer}, \& {Schad}}]{Morton_2025_ApJ}
{Morton}, R.~J., {Molnar}, M., {Cranmer}, S.~R., \& {Schad}, T.~A. 2025, \bibinfo{title}{{High-frequency Coronal Alfv{\'e}nic Waves Observed with DKIST/Cryo-NIRSP},} \apj, 982, 104, \dodoi{10.3847/1538-4357/adb8df}

\bibitem[{C.~J. {Nelson} {et~al.}(2024){Nelson}, {Hayes}, {M{\"u}ller}, {Musset}, {Freij}, {Auch{\`e}re}, {Aznar Cuadrado}, {Barczynski}, {Buchlin}, {Harra}, {Long}, {Parenti}, {Peter}, {Sch{\"u}hle}, {Smith}, {Teriaca}, {Verbeeck}, {Zhukov}, \& {Berghmans}}]{Nelson_2024_aa}
{Nelson}, C.~J., {Hayes}, L.~A., {M{\"u}ller}, D., {et~al.} 2024, \bibinfo{title}{{Spatial distributions of extreme-ultraviolet brightenings in the quiet Sun},} \aap, 692, A236, \dodoi{10.1051/0004-6361/202346886}

\bibitem[{L.~J. {November} {et~al.}(1981){November}, {Toomre}, {Gebbie}, \& {Simon}}]{November_1981_ApJ}
{November}, L.~J., {Toomre}, J., {Gebbie}, K.~B., \& {Simon}, G.~W. 1981, \bibinfo{title}{{The detection of mesogranulation on the sun.},} \apjl, 245, L123, \dodoi{10.1086/183539}

\bibitem[{N.~K. {Panesar} {et~al.}(2021){Panesar}, {Tiwari}, {Berghmans}, {Cheung}, {M{\"u}ller}, {Auchere}, \& {Zhukov}}]{Panesar_2021_ApJ}
{Panesar}, N.~K., {Tiwari}, S.~K., {Berghmans}, D., {et~al.} 2021, \bibinfo{title}{{The Magnetic Origin of Solar Campfires},} \apjl, 921, L20, \dodoi{10.3847/2041-8213/ac3007}

\bibitem[{S.~H. {Park} {et~al.}(2016){Park}, {Tsiropoula}, {Kontogiannis}, {Tziotziou}, {Scullion}, \& {Doyle}}]{Park_2016_AA}
{Park}, S.~H., {Tsiropoula}, G., {Kontogiannis}, I., {et~al.} 2016, \bibinfo{title}{{First simultaneous SST/CRISP and IRIS observations of a small-scale quiet Sun vortex},} \aap, 586, A25, \dodoi{10.1051/0004-6361/201527440}

\bibitem[{E.~N. {Parker}(1983){Parker}}]{Parker_1983_ApJ}
{Parker}, E.~N. 1983, \bibinfo{title}{{Magnetic Neutral Sheets in Evolving Fields - Part Two - Formation of the Solar Corona},} \apj, 264, 642, \dodoi{10.1086/160637}

\bibitem[{A. {Reiners}(2012){Reiners}}]{Reiners_2012_LRSP}
{Reiners}, A. 2012, \bibinfo{title}{{Observations of Cool-Star Magnetic Fields},} Living Reviews in Solar Physics, 9, 1, \dodoi{10.12942/lrsp-2012-1}

\bibitem[{M. {Rempel}(2017){Rempel}}]{Rempel_2017_ApJ}
{Rempel}, M. 2017, \bibinfo{title}{{Extension of the MURaM Radiative MHD Code for Coronal Simulations},} \apj, 834, 10, \dodoi{10.3847/1538-4357/834/1/10}

\bibitem[{F.~J. {Rogers} {et~al.}(1996){Rogers}, {Swenson}, \& {Iglesias}}]{Rogers_1996_ApJ}
{Rogers}, F.~J., {Swenson}, F.~J., \& {Iglesias}, C.~A. 1996, \bibinfo{title}{{OPAL Equation-of-State Tables for Astrophysical Applications},} \apj, 456, 902, \dodoi{10.1086/176705}

\bibitem[{S. {Shelyag} {et~al.}(2012){Shelyag}, {Mathioudakis}, \& {Keenan}}]{Shelyag_2012_ApJ}
{Shelyag}, S., {Mathioudakis}, M., \& {Keenan}, F.~P. 2012, \bibinfo{title}{{Mechanisms for MHD Poynting Flux Generation in Simulations of Solar Photospheric Magnetoconvection},} \apjl, 753, L22, \dodoi{10.1088/2041-8205/753/1/L22}

\bibitem[{M. {Shoda} {et~al.}(2023){Shoda}, {Cranmer}, \& {Toriumi}}]{Shoda_2023_ApJ}
{Shoda}, M., {Cranmer}, S.~R., \& {Toriumi}, S. 2023, \bibinfo{title}{{Formulating Mass-loss Rates for Sun-like Stars: A Hybrid Model Approach},} \apj, 957, 71, \dodoi{10.3847/1538-4357/acfa72}

\bibitem[{S.~S.~A. {Silva} {et~al.}(2024){Silva}, {Verth}, {Rempel}, {Ballai}, {Jafarzadeh}, \& {Fedun}}]{Silva_2024_ApJ}
{Silva}, S. S.~A., {Verth}, G., {Rempel}, E.~L., {et~al.} 2024, \bibinfo{title}{{Magnetohydrodynamic Poynting Flux Vortices in the Solar Atmosphere and Their Role in Concentrating Energy},} \apj, 963, 10, \dodoi{10.3847/1538-4357/ad1403}

\bibitem[{L. {Spitzer} \& R. {H{\"a}rm}(1953){Spitzer} \& {H{\"a}rm}}]{Spitzer_1953_PhRv}
{Spitzer}, L., \& {H{\"a}rm}, R. 1953, \bibinfo{title}{{Transport Phenomena in a Completely Ionized Gas},} Physical Review, 89, 977, \dodoi{10.1103/PhysRev.89.977}

\bibitem[{A.~C. {Sterling} \& R.~L. {Moore}(2020){Sterling} \& {Moore}}]{Sterling_2020_ApJ}
{Sterling}, A.~C., \& {Moore}, R.~L. 2020, \bibinfo{title}{{Coronal-jet-producing Minifilament Eruptions as a Possible Source of Parker Solar Probe Switchbacks},} \apjl, 896, L18, \dodoi{10.3847/2041-8213/ab96be}

\bibitem[{J.~O. {Thurgood} {et~al.}(2014){Thurgood}, {Morton}, \& {McLaughlin}}]{Thurgood_2014_ApJ}
{Thurgood}, J.~O., {Morton}, R.~J., \& {McLaughlin}, J.~A. 2014, \bibinfo{title}{{First Direct Measurements of Transverse Waves in Solar Polar Plumes Using SDO/AIA},} \apjl, 790, L2, \dodoi{10.1088/2041-8205/790/1/L2}

\bibitem[{V. {Upendran} {et~al.}(2022){Upendran}, {Tripathi}, {Mithun}, {Vadawale}, \& {Bhardwaj}}]{Upendran_2022_ApJ}
{Upendran}, V., {Tripathi}, D., {Mithun}, N.~P.~S., {Vadawale}, S., \& {Bhardwaj}, A. 2022, \bibinfo{title}{{Nanoflare Heating of the Solar Corona Observed in X-Rays},} \apjl, 940, L38, \dodoi{10.3847/2041-8213/aca078}

\bibitem[{S. {Wedemeyer-B{\"o}hm} \& L. {Rouppe van der Voort}(2009){Wedemeyer-B{\"o}hm} \& {Rouppe van der Voort}}]{Wedemeyer_2009_AA}
{Wedemeyer-B{\"o}hm}, S., \& {Rouppe van der Voort}, L. 2009, \bibinfo{title}{{Small-scale swirl events in the quiet Sun chromosphere},} \aap, 507, L9, \dodoi{10.1051/0004-6361/200913380}

\bibitem[{S. {Wedemeyer-B{\"o}hm} {et~al.}(2012){Wedemeyer-B{\"o}hm}, {Scullion}, {Steiner}, {Rouppe van der Voort}, {de La Cruz Rodriguez}, {Fedun}, \& {Erd{\'e}lyi}}]{Wedemeyer_2012_Natur}
{Wedemeyer-B{\"o}hm}, S., {Scullion}, E., {Steiner}, O., {et~al.} 2012, \bibinfo{title}{{Magnetic tornadoes as energy channels into the solar corona},} \nat, 486, 505, \dodoi{10.1038/nature11202}

\bibitem[{N. {Yadav} {et~al.}(2020){Yadav}, {Cameron}, \& {Solanki}}]{Yadav_2020_ApJ}
{Yadav}, N., {Cameron}, R.~H., \& {Solanki}, S.~K. 2020, \bibinfo{title}{{Simulations Show that Vortex Flows Could Heat the Chromosphere in Solar Plage},} \apjl, 894, L17, \dodoi{10.3847/2041-8213/ab8dc5}

\end{thebibliography}





\end{document}